\documentclass[sigconf]{acmart}
\AtBeginDocument{%
  \providecommand\BibTeX{{%
    \normalfont B\kern-0.5em{\scshape i\kern-0.25em b}\kern-0.8em\TeX}}}

\copyrightyear{2022}
\acmYear{2022}
\setcopyright{acmlicensed}\acmConference[ICSE-SEET '22]{44nd International Conference on Software Engineering: Software Engineering Education and Training}{May 21--29, 2022}{Pittsburgh, PA, USA}
\acmBooktitle{44nd International Conference on Software Engineering: Software Engineering Education and Training (ICSE-SEET '22), May 21--29, 2022, Pittsburgh, PA, USA}
\acmPrice{15.00}
\acmDOI{10.1145/3510456.3514159}
\acmISBN{978-1-4503-9225-9/22/05}

\usepackage{pgfplots}
\pgfplotsset{compat=1.17}

\begin{document}

\title[Pattern Matching and String Completion for Teaching Programming]{Write a Line: Tests with Answer Templates and String Completion Hints for Self-Learning in a CS1 Course}

\author{Oleg Sychev}
\affiliation{%
  \institution{Volgograd State Technical University}
  \streetaddress{Lenin Ave, 28}
  \city{Volgograd}
  \country{Russia}
}
\email{oasychev@gmail.com}
\orcid{0000-0002-7296-2538}

%\author{Artem Prokudin}
%\affiliation{%
%  \institution{Volgograd State Technical University}
%  \streetaddress{Lenin Ave, 28}
% \city{Volgograd}
%  \country{Russia}
%}
%\email{prokudintema@yandex.ru}
%\orcid{0000-0002-0694-0808}

%\renewcommand{\shortauthors}{Sychev O. and Prokudin A.}
\renewcommand{\shortauthors}{Sychev O.}

\begin{abstract}
  One of the important scaffolding tasks in programming learning is writing a line of code performing the necessary action. This allows students to practice skills in a playground with instant feedback before writing more complex programs and increases their proficiency when solving programming problems. However, answers in the form of program code have high variability. Among the possible approaches to grading and providing feedback, we chose template matching. This paper reports the results of using regular-expression-based questions with string completion hints in a CS1 course for 4 years with 497 students. The evaluation results show that Perl-compatible regular expressions provide good precision and recall (more than 99\%) when used for questions requiring writing a single line of code while being able to provide string-completion feedback regardless of how wrong the initial student's answer is. After introducing formative quizzes with string-completion hints to the course, the number of questions that teachers and teaching assistants received about questions in the formative quizzes dropped considerably: most of the training question attempts resulted in finding the correct answer without help from the teaching staff. However, some of the students use formative quizzes just to learn correct answers without actually trying to answer the questions.
  %Добавить про активность использования с ростом курса.
\end{abstract}





\maketitle

\section{Introduction}

Introducing students to the programming domain is a complex process, requiring the development of a significant number of cognitive skills \cite{Luxton2018}. It spans all levels of revised Bloom's taxonomy of educational objectives \cite{AndersonKrathwohl2001} from "remember" to "create", having strict constraints of logical correctness on all the levels to get a correct solution. Different e-learning tools may be used to support students while mastering different kinds of learning tasks.

One of these tasks is writing simple statements or headers, learning both syntax and semantics of the programming language \cite{Chirumamilla2019EAssessment}. These exercises belong to the application level of Bloom's taxonomy because they require applying syntax rules in typical situations. They lay the foundation for higher-level tasks such as analysis of existing programs and synthesis of new programs. While some of the students can write programs directly after being introduced to a text-based programming language, many others need scaffolding to get used to the syntax before attempting more complex tasks. Automated "write a line of code" assessments give them a convenient playground to acquire and reinforce basic skills. Without good skills in writing single lines of code, students make a lot of mistakes in programming assessments when they are focused on higher-level planning of the algorithm.

E-assessment is a good solution for the exercises requiring writing a single line of code because grading the required amount of questions manually requires too much time. However, simple short-answer question software is poorly suited for this situation because the response in the form of a line of code has high variability; many programming-language features can increase the set of correct answers -- from the ability to insert any number of white space characters almost in any place allowing them without changing the answer -- to the problem of extraneous (but correctly opened and closed) parentheses in expressions. Without a way to define the set of correct answers using templates or other mechanisms, code-writing short-answer questions become restrictive, giving a lot of false-negative grades and frustrating students.

When code-writing questions are used for formative assessments, it is necessary to provide feedback to the students about their mistakes and the ways to correct them. Otherwise, the students become stuck, requiring intervention from the teacher to understand how to complete their current exercise. Ott \cite{Ott2016} supposes using targeted questions and hints to guide students in the process of finding the answer to a problem. Automatically generated feedback allows using formative assessments as homework with a significantly higher amount of completed questions than while utilizing teacher-provided feedback.

Some authors argue for parsing and executing program code to assess it (e.g., \cite{AlaMutka2005}), a template-based approach can be useful for smaller problems involving writing a few lines of code. When the student's task is to write a single line of code, executing it often requires a significant number of enclosing code that must be supplied by the teacher, and testing errors often correspond poorly with the original task. 

It is possible to parse students' code and compare resulting syntax trees, but this method also has significant disadvantages. First, there is no single starting symbol of the grammar for parsing a single line of code - the starting symbol depends on the particular question. Sometimes, a line of code cannot be reduced to a single symbol at all. Also, parsing does little to account for variation in the programming language (for example, in the C++ language, there are 6 ways to add 1 to a variable) so the teacher still has to provide all correct solutions. This is especially problematic if the answer contains several parts that can be written differently, as the number of correct answers rises multiplicatively. Template-based systems can solve this problem by letting to describe the variants of each part separately, significantly decreasing the effort required to create a question. Error-correcting parsers are also computationally expensive \cite{rajasekaran2014error} while pattern matching can be implemented efficiently using, for example, finite automata.

%%More links
The most important advantage of string templates over parsing is that they allow easy generation of completion hints. While there are works on generating data-driven feedback based on comparing syntax trees \cite{Marin2021}, they only work for fairly close syntax trees. However, sometimes a single wrong character can lead to a significantly different syntax tree. Also, as we often saw in teaching practice, some poorly-performing students who need hints most give answers that are very far from the expected line of code while error-correcting parsers are only good if there are a few errors. Errors of this kind happen more often in the first stages of learning programming (or when moving from a block-based programming language to a text-based language) where the questions requiring writing a single line of code are used. Novice programmers also rarely think about their programs in the terms that are expressed in the language grammars. So for short answers, comparing string representation of the program code may yield better hints than comparing syntax trees for the courses introducing students to the text-based programming language.

A commonly-used way to specify string-matching templates is regular expressions; they are expressive and relatively compact. Regular expressions were used in automated programming assessment to verify the output of the tested program \cite{Higgins2005}; they were also used for implementing short-answer questions \cite{butcher2010comparison}. However, most question-building tools do not allow combining advanced regular-expression syntax with string completion hints.

In this paper, we describe our experience of introducing formative quizzes with string completion hints into a CS1 course of Volgograd State Technical University, analyzing the results of four years of teaching. The quizzes were created using the Preg question type for Moodle LMS: a short-answer question software using Perl-compatible regular expressions to define correct answers and providing advanced feedback, showing partial matches (i.e., the correct beginning of the student's answer up to the first error) and providing string-completion hints (next correct character and next correct lexeme). The rest of the paper is organized as follows. Section 2 describes the state of the art in template-based assessments and string completion hints; section 3 provides the background in regular expressions; section 4 describes the developed software; section 5 shows our findings while analyzing our experience followed by discussion in section 6 and conclusion in section 7.

\section{Related Work}

\subsection{Template-based short-answer assessment}

Existing systems and approaches use different ways to specify patterns for short answers \cite{Burrows2015ErasAutomaticAnswerGrading,SYCHEV2020264}. Some of those systems calculate grades by performing a word-by-word comparison of weighted words, other systems use specific languages to describe patterns.

WebLAS identifies important segments of correct answers in parsed representations and asks the teacher to confirm each of them and assign weights to them \cite{Bachman2002AutomaticAssessmentShortAnswer}. The teacher is also asked to accept or reject semantically similar alternatives. Each segment of the student's answer is detected using regular expressions. Each segment is graded separately, so it is possible to grade partially-correct answers.

eMax uses a similar approach - it requires the teacher to markup semantic elements \cite{Sima2007IntelligentShortTextAssessment}. Teachers can also accept or reject synonyms and assign weights to each answer element. The grading approach is combinatorial, i.e. all possible formulations are pattern-matched. The assigned scores can be forwarded for manual review in difficult cases.

FreeText Author requires teachers to describe answers with syntactic-semantic templates for the student answers to be matched against \cite{Jordan2009PotentialShortAnswer}. Such templates are automatically generated from the plain-text representation of the teachers' answers. Through the interface, the teacher can specify mandatory keywords from the correct answers and select synonyms provided by thesauri support. Both acceptable and unacceptable answers can be defined. 

IndusMarker uses word- and phrase-level pattern matching to grade student answers \cite{Siddiqi2008AutomatedMarkingShA}. Credit-worthy phrases are defined using an XML-based markup language called the Question Answer Markup Language. Using the “structure editor”, the text and number of points can be specified for each phrase. 

OpenMark uses its own template language \cite{butcher2010comparison}. It can be used for specifying alternative words (e.g., synonyms) or word groups. The system also supports typo detection.

Most of these systems are aimed specifically at grading natural-language answers and are poorly suited for programming-language answers.

%CorrectWriting plugin for Moodle LMS uses combined character- and word-level analysis to grade answers and provide feedback to students\cite{sychev2018automatic}. Word-level analysis uses the longest common sequence (LCS) algorithm \cite{apostolico1997string}, whereas character-level analysis uses word comparison with edit distance for typo detection. 

\subsection{String-completion hints}

Providing detailed feedback is an important requirement for short-answer grading systems. When a student makes a mistake, it is not enough to simply tell them that the answer is wrong \cite{Hattie2007PowerFeedback, gibbs2005conditions, BlackWiliamAssessmentLearning1998}. They still have to either read their lecture materials to identify their mistakes or wait for the opportunity to consult with their teacher. Falkner et al. \cite{Falkner2014} suggested that increasing feedback granularity impacts performance positively. Automatic feedback, including the information about possible mistakes and the ways to fix them, significantly increases the efficiency of student efforts, allowing them to understand and correct their mistakes during an online session at a convenient time for them.

There are several well-known automatic feedback generation techniques \cite{Keuning2016SystematicReviewAutomatedFeedback}. 

\begin{itemize}
    \item Intelligent Tutoring System model-tracing feedback generation technique that uses comparison of student's steps with buggy production rules \cite{Corbett2001Feedback, Gerdes2012FunctionalTutor}.
    \item Constraint-based feedback generation technique that checks student's answer against predefined constraints and generates feedback messages for failed ones \cite{Nguyen2006LogicProgramming}. 
    \item Natural Language Processing and Machine Learning feedback generation technique that provides feedback using datasets of students' answers \cite{Lane2005NLTutoring, Macnish2010AutomatingFeedback, Dzikovska2014}.
\end{itemize}

%AutoGrader is testing system for programs written in python. It can identify and correct errors when the wrong code is sent, and then automatically generate sequences of increasingly specific clues about where the error is and what characters need to be changed to fix it \cite{SinghRishabhAutomatedFeedback2013}.

\section{Background}

Regular expressions are a standard method of specifying string patterns \cite{friedl2006mastering}. Applied to teaching programming, one of their undoubted advantages is their versatility regarding answer languages. They allow creating complex patterns that cover wide ranges of possible correct answers. Many tutorials and software tools for creating and debugging regular expressions are available \cite{BeckFabian2014RegViz, Budiselic2007RegExpert}, but let us explain the very basics for the reader to be able to continue reading on.

Regular expressions, just like any other expressions, consist of operators and operands. The simplest operand is a Latin character: it matches itself. There are also character sets written like \verb%[0-9a-fA-F]% (matches hexadecimal digits), meta-characters (a dot matches any character except line breaks), escape-sequences (\verb%\s% matches whitespaces and \verb%\d% matches decimal digits), and so on.

Operands can be \textit{concatenated} to match a sequence of characters: \verb%\d\s\d% matches any two decimal digits with whitespace between them. They also can be \textit{quantified} to match a sequence of repeated strings: \verb%(\d\s)+% matches a digit followed by a whitespace, repeated any number of times (but at least once). The \verb%*% matches zero or more repetitions of its operand, and the \verb%?% matches 0 or 1 repetitions. Finally, operands can be \textit{alternated}: \verb%ab|cd+% matches either ``ab'' or ``cd'', ``cdd'', ``cddd'' and so on. Note that concatenation takes precedence over alternation, and quantification takes precedence over concatenation.

Perl-compatible regular expressions (PCRE) have the richest syntax and most powerful features \cite{goyvaerts2012regular}. In addition to the above-mentioned features, they support such powerful features as named subpatterns, backreferences, recursive subpattern calls, look-around assertions, lazy and possessive quantifiers, extended syntax for character classes, and others. In the following sections, regular expressions will mean Perl-compatible regular expressions.

\section {Preg Question's Features and Their Uses in Introductory Programming Courses}

\subsection{Regular Expressions as Templates for Program Strings}

\subsubsection{Basic Regex Features}

The first problem to solve when checking a student's program string is accepting any correct whitespace placement: most of the programming languages allow any number of whitespace characters between any lexemes and require at least one whitespace between some.%For example, in just a simple integer variable declaration in C/C++ there might be any number of whitespaces between the type and the variable's name, as well as before the trailing semicolon. In this case, we need to use regex quantifiers to match whitespace repetitions, and the correct pattern would look like \verb%int\s+([_a-zA-Z]\w*)\s*;% where the parenthesized part matches any valid variable name just like C/C++ compilers do.
We can use repetition of\verb|\s| character class to specify this e.g., \verb%int\s+([_a-zA-Z]\w*)\s*;%

Another useful regular expression feature is alternation. If a student was asked to write a floating-point variable declaration, both ``float'' and ``double'' types should be accepted, so our previous example would turn into this: \verb%(float|double)\s+[_a-zA-Z]\w*\s*;%. While this can be solved without regular expressions by providing a set of correct answers, their number increases multiplicatively when the answer contains several sections with independent alternatives and quickly becomes impractical. 

\subsubsection{Advanced Regex Features}
\label{adv-features}
One more useful feature of regular expressions is back-referencing. A back-reference works in pair with a subpattern (a parenthesized part of the regular expression) and matches exactly the same string as its corresponding subpattern does. Assume that a student needs to write a ``for'' loop with an integer counter running from 0 to 9 and an empty body in the C or C++ languages. The counter's name should appear in the answer 3 times unmodified, and this can be achieved by enclosing the variable name in parentheses (thus creating a subpattern) and then referencing it two times later in the expression like this:\linebreak\verb%for\(int ([_a-zA-Z]\w*)=0;\1(<=9|<10);\1\+\+\)(;|\{\})% where \verb|\1| is a backreference to the first subpattern (i.e., the expression inside the first set of unescaped parentheses counting opening parentheses from the left).  Correct whitespace matching is ignored here for the purpose of readability. This pattern matches student answers like ``for (int i=0;i<=9;i++);'' and ``for (int myVar=0;myVar<10;myVar++)\{\}''.

One of the biggest challenges with grading students' answers containing formulas and program code using string templates is matching an arbitrary but correct placement of parentheses around expressions, i.e., ``5'' and ``(((5)))'' should be graded the same. Regular expressions using recursive subpattern calls (an exclusive feature of Perl-compatible regular expressions) can solve this problem, though the expressions become a lot harder to understand. Here is a regular expression matching the parenthesized digit 5: \verb%5|(\s*\(\s*(?:(?-1)|5)\s*\)\s*)%.
%% \begin{math}5|(\textbackslash{} s*\textbackslash{}(\textbackslash{}s*(?:(?-1)|5)\textbackslash{}s*\textbackslash{})\textbackslash{}s*)\end{math}

It is clear that recursive regular expressions cannot be used for real-life questions ``as is'', because usually there are several subexpressions that need to be parenthesised so the whole regular expression becomes less and less readable. To solve this problem, we introduced a set of special regex comments in Preg which look like the following:
\begin{itemize}
    \item \verb%(?###parens\_opt<)body(?###>)%;
    \item \verb%(?###parens\_req<)body(?###>)%.
\end{itemize}
These expressions define patterns match everything that their ``body'' matches, but is enclosed in zero or more(\verb|parens_opt|) or one or more (\verb|parens_req|) parentheses. There is also a number of other predefined patterns for matching brackets, custom parentheses (with any opening/closing symbols), identifiers, etc. With this feature, strings like ``5'' and ``(((5)))'' can be matched by a simple and readable regular expression: \verb%(?###parens_opt<)5(?###>)%.% These recursive expressions turned out to be advantageous as shown later in section~\ref{s:answer_templates}.

\begin{figure}[h]
  \centering
  \includegraphics[width=\linewidth]{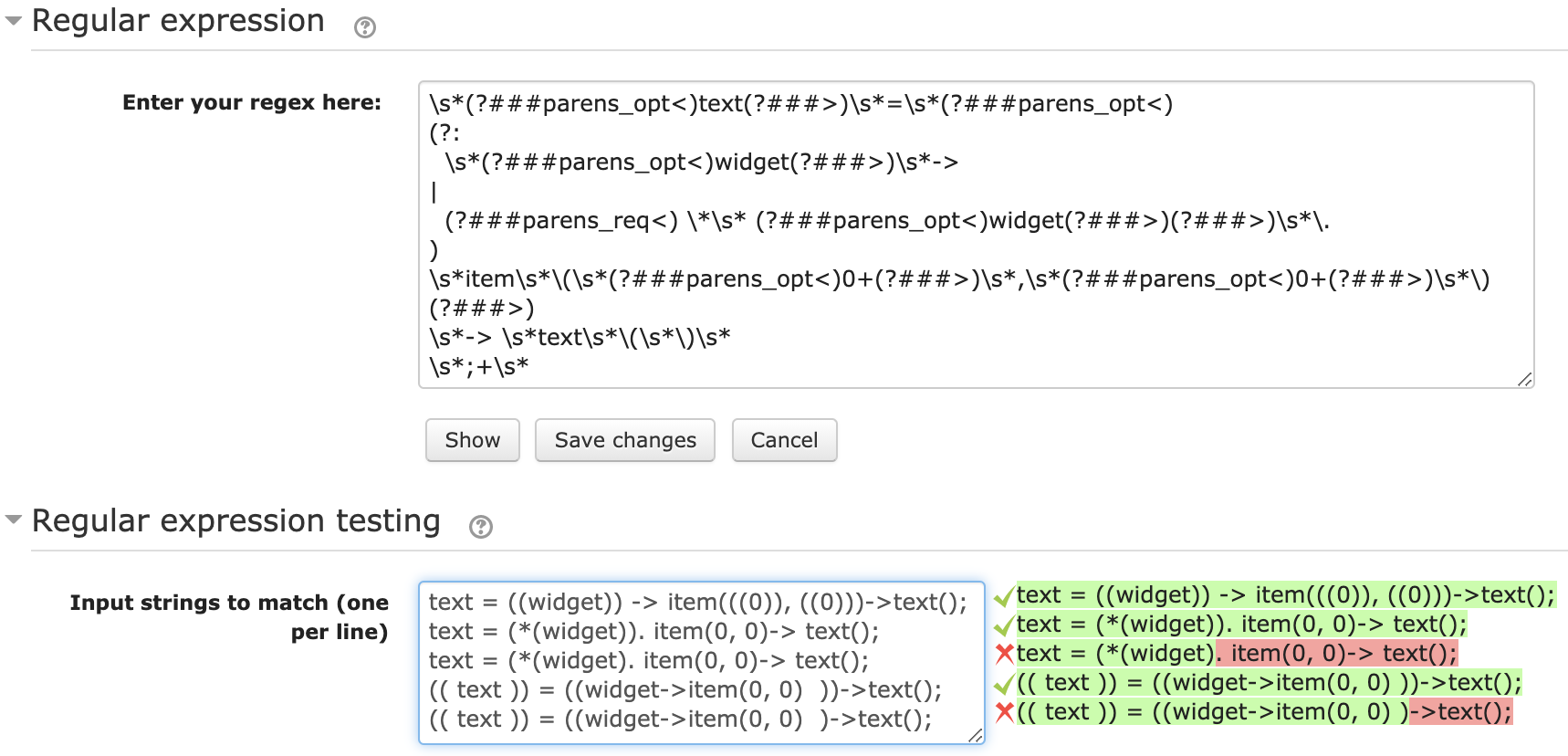}
  \caption{The tool for editing and debugging regular expressions in Preg questions\label{fig:feature_matching}}
  \Description{There is a multi-line text input field with a complex regular expression. Under that field, there is another one with 5 answers entered line-by-line. The answers mostly vary in parentheses placement. Three of them are fully-matched and highlighted green, while the other two are matched partially and have the incorrect parts highlighted red. There are green check- and red cross-marks next to each of the answers.}
\end{figure}

In Figure~\ref{fig:feature_matching} a regular expression from an actual code writing question is shown. The expression is entered in a multi-line text field and is still readable because of using the comment-based patterns. Below is a testing tool for debugging the expression, showing the range of possible correct answers these expressions match. The tool highlights the matched and unmatched parts of the answers.

\subsubsection{Partial Matching}
Unlike the commonly-used regex libraries, the Preg question's engine supports finding partial matches, i.e., matches starting at the beginning of the regex and breaking off somewhere in the middle. With this feature, on the higher level Preg highlights correct and incorrect parts of students' answers in green and red, respectively. 

\subsection{String Completion Hints}
The Preg question type is capable of generating completions of partial matches, leading to full matches. The completion is always chosen so that they will lead the student to the correct answer adding the minimum number of characters. Based on this, Preg can show two hints: the next correct character and the next correct lexeme. Hints may be turned off (for summative quizzes); a penalty score can be specified for each hint usage to discourage students from overusing the hints.

\begin{figure}[h]
  \centering
  \includegraphics[width=\linewidth]{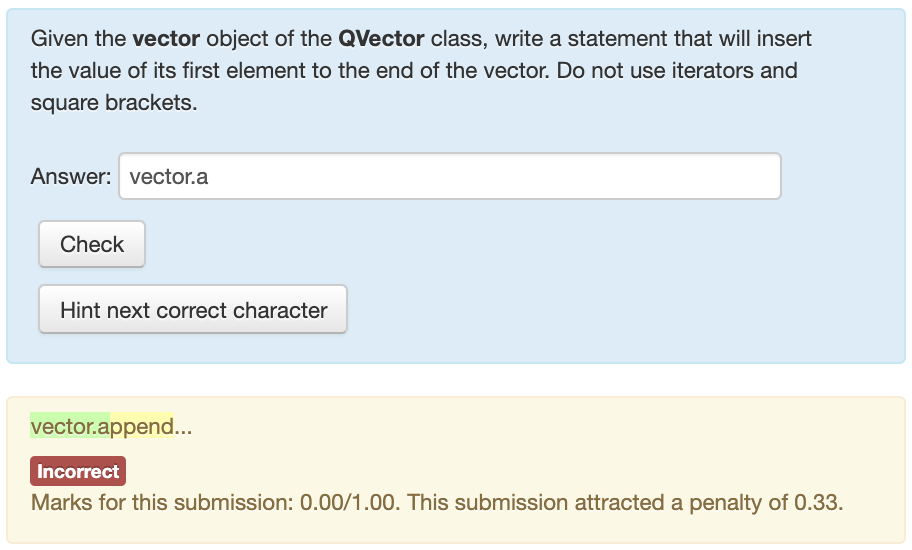}
  \caption{Next correct lexeme hint in a Preg question\label{fig:feature_hinting}}
  \Description{A window containing the question text, a text input field with a partially entered answer highlighted green, and a next-lexeme hint highlighted yellow. The hint is followed by an ellipsis indicating that more text should be written to complete the answer.}
\end{figure}

Figure~\ref{fig:feature_hinting} demonstrates how the next correct lexeme hint looks in Preg questions. Here, a student asks to complete the name of the method to call; the hinted part is highlighted yellow. The ellipsis after it shows the student that the answer is not complete after adding it.

%\subsection{Special patterns for catching common mistakes}
%On a higher level, it is possible to define separate answers (patterns) matching the most common mistakes that students do. In the ``integer variable declaration'' example we could specify a pattern matching answers without trailing semicolon. Usually such patterns give a 0-points grade and show detailed feedback to the student.

\section{Case Study}

Volgograd State Technical University uses Preg questions in the two-semester CS1 course ``Programming basics'' for the first- and second-year undergraduate students. The following data was gathered from the first semester of this course (first year, spring semester) unless explicitly state otherwise because the first semester had more academic hours and so more quizzes; the first semester of this course is also focused introducing students to the syntax of the C++ programming language that is better suited for the questions requiring writing a line of code.

\subsection{Course Quiz Setup}

The first semester of the CS1 course ``Programming basics'' contains 6 major programming assignments for topics ranging from alternative statements to user-defined data types. Each of them has an associated summative quiz (10 questions, about 15 variants each) that the student must pass in class before attempting the assignment. These quizzes help to make sure that the student knows the topic enough to attempt writing a program. Hints are disabled in summative quizzes, and students cannot re-attempt summative questions once they are graded. The questions in summative quizzes include code-writing Preg questions (from 1 to 5 questions per quiz), multiple-choice questions ``find the code lines containing errors'', and short-answer questions ``determine the results of code execution''. 

Originally, to prepare students for the summative quizzes, for each summative quiz there was created a regular formative quiz, offering a few variants of the difficult questions, and a demonstration quiz with one variant of each question of the summative quiz. All the questions in these formative quizzes were supplied with teacher-defined natural-language feedback, explaining which answer is correct and why. The students actively used the formative quizzes, however, the natural language explanatory feedback was not enough: each lab assignment started from answering students' questions about the formative quizzes they were stuck at which reduced the time for actual programming. The main problem for the code-writing questions was that the basic Moodle short-answer questions just reported the fact of error; the students had to choose between trying to fix their error by guessing and learning the correct answer.

To enhance the students' abilities to train during homework, we developed Preg formative quizzes for each code-writing question of each summative quiz with string-completion hints enabled. Students can use formative quizzes as much as they want during homework but have no obligation to attempt them. Each formative quiz usually has about 5 variants of each question. While attempting a formative quiz, the student can correct their answer and re-grade it as many times as they want. They also can use the string-completion hints that the Preg question type provides.

The summative quizzes are graded, but their influence on the course grade is minimal (less than 10\% of the course grade). Their main function is being the threshold to cross before attempting more complex tasks. This was done to limit the grade loss caused by attention slips during testing as the students cannot debug their code while attempting summative quizzes. The formative quizzes are provided for training purposes and do not affect the course grade. All quizzes showed the correct answers and natural-language feedback once the quiz is completed.

The students who fail a summative quiz (by scoring less than 60\%) are given time to improve their skills and should re-attempt the quiz until passing it (up to 5 attempts). They can use formative quizzes while preparing for the second and later attempts. As a part of the final exam, students pass the exam quiz containing 18 questions from all 6 summative quizzes, providing a broad overview of their knowledge of the whole course.

After introducing Preg formative quizzes, the main students' complaint about them was its lack of handling of extra parentheses in expressions which produced false-negative grades. To resolve this problem, we enhanced the existing regular expressions with recursive subpattern calls using the special Preg syntax described in the section~\ref{adv-features}. It resolved the problem, significantly lessening the number of false-negative grades.

%\subsection{Method}
The results presented below are based on the data collected over four years (2016, 2017, 2019, and 2020) of teaching an introductory programming course in Volgograd State Technical University to the first-year students specializing in ``Informatics and Computing'' and ``Software Engineering'' with 479 students in total.

%The students had to pass 6 summative quizzes (for each topic) and an exam quiz at the end of the semester; they were allowed to use formative quizzes with Preg questions freely to prepare for the summative and exam quizzes. %As automatically-generated feedback is most useful for self-directed learning during preparation for classes that cannot be controlled by the tutoring staff, we aimed at evaluating students' natural behavior when additional exercises with feedback are available and did not interfere with their choice of using the formative quizzes; their only motivation was improving their knowledge and preparing for the exam better.
We gathered all the data about the question attempts, including their sequence, students' responses to each question, and the hints they requested. We also conducted informal interviews with the staff teaching the course.% This allowed us to evaluate both the templating abilities of the Preg question type and the usage of hint feature by the students.
%%%Надо ли?
%Some of the students completed a short survey about using Preg questions in their learning.

\subsection{Answer templates}\label{s:answer_templates}
%ed
The course contains 722 Preg questions using 1160 regular expressions in total. One Preg question contained from 1 to 13 regular expressions (mean 1.5 regular expressions per question) used to test various correct answers. These regular expressions described program codes of varying lengths that can be measured in characters and tokens. In Table~\ref{table-regex-measures} we provided measures for the shortest correct answer and the length of the longest correct answers for most of the regular expressions is (theoretically) not limited because of quantifiers ``*'' and ``+'' allowing any number of whitespace characters. Another important measure of the regular expression complexity is the number of possible alternative paths through the expressions, showing the variability of code described by them.

\begin{table}
  \caption{Quantitative analysis of the regular expressions used in ``Programming Basics'' course}
  \label{table-regex-measures}
  \begin{tabular}{lcccc}
    \toprule
    Parameter&Min&Max&Mean&Stdev\\
    \midrule
    characters in answer & 2 & 109 & 31.08 & 19.8\\
    tokens in answer & 1 & 70 & 14.22 & 8.75\\
    paths through the expression & 1 & 18 & 1.58 & 1.86 \\
  \bottomrule
\end{tabular}
\end{table}
%ed

The students gave 4952 unique answers that were matched by these regular expressions between 1 to 67 unique answers per regular expression, 4.9 unique answers per expression on average.

\begin{figure*}[h]
   \centering
  \resizebox{0.8\textwidth}{!}{

   \begin{tikzpicture}
    \begin{axis}[
        ybar,
        bar width=2pt,
        xlabel={number of unique answers},
        ylabel={number of regular expressions},
        xmin=1, xmax=30,
        ymin=0, ymax=750,
        xtick={0,2,4,6,8,10,15,20,25,30},
        %ytick={0,20,40,60,80,100,120},
        %legend pos=north west,
        ymajorgrids=true,
        grid style=dashed,
    ]
    \addplot[
        %smooth,
        color=blue,
        fill=cyan,
        ]
        coordinates {
        (1, 302)(2, 297)(3, 246)(4, 206)(5, 163)(6, 129)(7, 103)(8, 78)(9, 63)(10, 55)(11, 41)(12, 32)(13, 26)(14, 23)(15, 17)(16, 14)(17, 10)(18, 9)(19, 7)(20, 6)(21, 5)(22, 4)(23, 4)(24, 4)(25, 3)(26, 2)(27, 2)(28, 2)(29, 1)
        };
    \addlegendentry{unique answers}
    \addplot[
        %smooth,
        color=orange,
        fill=orange,
        ]
        coordinates {
        (1, 750)(2, 599)(3, 431)(4, 311)(5, 224)(6, 162)(7, 117)(8, 86)(9, 62)(10, 44)(11, 32)(12, 24)(13, 17)(14, 13)(15, 10)(16, 8)(17, 5)(18, 4)(19, 3)(20, 2)(21, 2)(22, 2)(23, 1)(24, 1)(25, 1)(26, 1)(27, 1)(28, 1)(29, 1)
        };
    \addlegendentry{suppressing difference in white spaces}    
        
    \end{axis}
    \end{tikzpicture}

%    \includesvg{unique_correct_answers}
    }

  \caption{Distribution of regular expressions by the number of unique covered answers}
    \label{regex-unique-answers-distribution}
  \Description{The chart contains two curves. Both of them decline steadily, with the ``unique answers'' one starting at the 2-times lower point.}
\end{figure*}
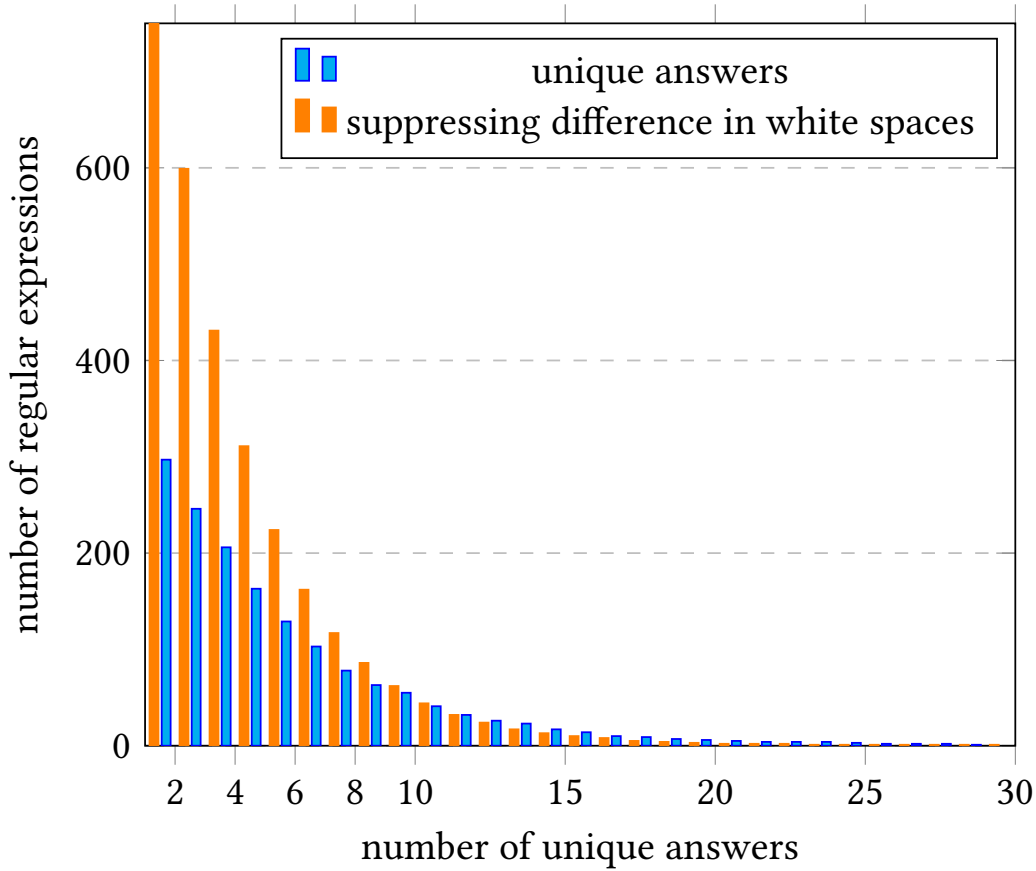

Some of these unique answers differed only in whitespace characters. Considering all answers that differ only in whitespaces the same, we found 2107 unique answers, between 1 to 40 answers per regular expression, 2 answers per expression on average. The distribution of the number of unique answers per regular expression is shown in Fig.~\ref{regex-unique-answers-distribution}: the number of unique answers on the X-axis, and the number of regular expressions with this number of unique answers on the Y-axis. The majority of regular expressions had from 1 to 5 unique answers; the number of expressions matching more than 15 unique answers is small. It shows that using regular expressions allowed us to cover a wider range of correct answers than it would be possible using simple open-answer questions with string matching.

%% These were the most... переформулировать
%As was said above, one feature of Perl-compatible regular expressions that is particularly useful when teaching computer science is recursive subpattern calls: this feature allows to solve the problem of extraneous parentheses in expressions. These were the most often false negative grades that the teachers in the Introductory Programming course encountered after using regular expressions for testing. Few regular expression engines support recursive subpattern calls, and none of them generates partial match completions for hinting.

%After implementing recursive subpattern calls and special template syntax to make them readable, we upgraded all the questions in the course that could contain parentheses and measured quiz system performance. 
As we said before, the main cause of false grades for regular expressions questions were extraneous parentheses.
The usage of recursive subpattern calls to catch extraneous parentheses increased recall for the Preg question in the entire course from 0.88 to 0.99 and F-measure from 0.937 to 0.995. The few remaining cases of false-negative grading are solved by teachers individually; sometimes regular expressions were upgraded to match a new solution found by a student.

\subsection{Formative Quizzes and Hints}
%Evaluating the developed tool, we let the students choose freely how much they will use formative Preg quizzes among the other ways to prepare for the summative quizzes (including regular and CorrectWriting formative quizzes).
When evaluating Preg formative quizzes' effect on the students' performance, it must be taken into account that Preg quizzes were working with only one kind of question - open-answer string questions - which comprised roughly one-third of the summative quizzes. About 43\% of the students chose not to use Preg quizzes at all. 
%%%New_2_1
%To evaluate the usage of the developed tool, we analyzed how the students used the formative quizzes.
To understand the next two tables better, it is good to keep in mind the difference between the number of quiz attempts and the number of question attempts: each formative quiz contains from 1 to 5 questions depending on the topic.

Out of 464 students who attempted at least one quiz in the course, 207 (44.6\%) chose to not use Preg formative quizzes at all. The remaining 257 students used Preg quizzes for training from 1 to 107 times (mean 9.16, standard deviation 13.98). They made 3100 quiz attempts. Table~\ref{table-2-quiz-attempts} summarizes the ways the students used Preg formative quizzes. Given that they could improve their answers until answering correctly or giving up, we can consider correct answers as the student being able to complete their attempt (or some questions in it).

\begin{table}
  \caption{Attempts of formative quizzes}
  \label{table-2-quiz-attempts}
  \begin{tabular}{cccc}
    \toprule
    Tried to answer&Answered&Number of&Percentage\\
          questions&correctly&attempts& \\
    \midrule
    No & No & 87 & 2.8\% \\
    Some & No & 76 & 2.5\% \\
    Some & Some & 239 & 7.7\% \\
    All & Some & 738 & 23.8\% \\
    All & All & 1960 & 63.2\% \\
  \bottomrule
\end{tabular}
\end{table}

It can be seen see that more than half of the training attempts ended in answering all questions correctly. In more than three-quarters of attempts, the student answered correctly at least some questions. Only in 5.3\% of attempts the student gave up without answering at least one question correctly. Half of them simply used formative quizzes to learn the correct answers, clicking the button to complete the quiz right after it started.

%%%New_2_2

Looking at how the individual questions were attempted, we found 8291 question attempts. The summary is shown in Table~\ref{table-2-question-attempts}. Note that for individual questions, the number of attempts without even trying to answer should be interpreted differently: it is a valid strategy to answer only some questions during a quiz attempt meant for training if the student is sure in their ability to answer the other questions. The students were able to find correct answers to almost 80\% of the questions without support from the teaching staff. About one-third of the completed attempts used hints. In 13.6\% of attempts, the student gave up solving a question. In 27\% of these attempts, the student used hints but still did not find the correct answer. The percentage of not completed attempts is almost the same for the attempts with hints (13\%) and the attempts that did not use hints (15\%).

\begin{table}
  \caption{Attempts of formative questions}
  \label{table-2-question-attempts}
  \begin{tabular}{cccc}
    \toprule
    Final&Number of&Percentage&Attempts with\\
    answer&attempts& &hints\\
    \midrule
    Absent & 581 & 7\% & 0 \\
    Incorrect & 1126 & 13.6\% & 304 \\
    Correct & 6584 & 79.4\% & 2000 \\
  \bottomrule
\end{tabular}
\end{table}

To assess the effect of using Preg quizzes on the overall performance of the students, we studied average exam quiz scores for different groups of students. The exam quiz was not obligatory, and some students were not allowed to take exams because of not completing enough assignments during the semester. We found 272 attempts of the exam quiz. Its results depending on the number of attempts of Preg formative quizzes are shown in Table~\ref{table-2-5-att}, while depending on the frequency of hint usage are shown in Table~\ref{table-2-5-hints}. Paired, 2-tailed t-test does not show any significant difference in the exam quiz scores according to the number of attempts of Preg formative quizzes. However, the students who used hints sparingly performed significantly better than those who did not use hints at all (p=0.02) and those who used hints often (p=0.05).
It is clear that no significant influence of Preg training on the exam quiz can be found. This may be unsurprising, given that the students had other means of training for the exam quiz and other means to get help. When we consider hint usage, the small group of students who used hints sparingly showed marginally better performance than the students who did not use hints or used them often. However, the small size of this group and the weak effect size require further research to validate this finding. The interviews with the teaching staff indicated that the main advantage of introducing Preg formative quizzes was significantly reducing the number of questions about homework formative quizzes they had to answer as the students, with the help of string-completion hints, could resolve many problems on their own.

\begin{table}
  \caption{The exam quiz score by the activity in using formative quizzes during the semester}
  \label{table-2-5-att}
  \begin{tabular}{cccc}
    \toprule
    Student Group&Students&Mean&Stdev\\
    \midrule
    no Preg training & 99 & 0.73 & 0.16\\
    low Preg training (<= 7 attempts ) & 114 & 0.729 & 0.16\\
    active Preg training (> 7 attempts) & 59 & 0.76 & 0.16\\
  \bottomrule
\end{tabular}
\end{table}

\begin{table}
  \caption{The exam quiz score by the percent of hints used during the semester}
  \label{table-2-5-hints}
  \begin{tabular}{cccc}
    \toprule
    Student Group&Students&Mean&Stdev\\
    \midrule
    did not use hints & 142 & 0.72 & 0.15\\
    low hint usage & 30 & 0.78 & 0.13\\
    (< 33\% of attempts used hints) & & & \\
    active hint usage & 100 & 0.73 & 0.16\\
    (> 33\% of attempts used hints) & & &\\
  \bottomrule
\end{tabular}
\end{table}

\section{Discussion}

While automatic feedback generation for programming exercises was researched extensively (see \cite{Keuning2018FeedbackProgrammingExercises}), the effects of smaller-size tasks like writing a single line of code did not receive the same attention. The reports on the effects of voluntary programming practice in introductory programming courses differ. Spacco et al \cite{Spacco2015} reported a weak correlation between exam performance and the number of voluntarily completed programming exercises. Estey et al \cite{Estey2017} reported a negative correlation between the number of requested hints and exam performance while no relationship between practice time and exam performance. Edwards et al \cite{Edwards2019} divided students into three groups (no practice, some practice, and full practice) and found that students from the ``some practice'' groups performed worst in both code-writing exercises and multiple-choice questions in final exams. However, all this research regarded voluntary solving program-writing exercises.

%In our study, we evaluated the usage of our tool in an introductory programming course and the role of string-completion hints in the preparation for final exams.

%\textbf{Q1. How well regular expressions are suited for open-answer questions in an introductory programming course?}
We found that regular expressions are suited well for programming questions with short answers (about one line of code; from 1 to 70 tokens) which are used in introductory programming courses during quizzing. Regular expressions allowed teachers to create questions with up to 18 significantly different answers (i.e., the answers that are different in more than just extra whitespaces and parentheses) per expression which would be time-consuming if using string matching or parse trees matching. However, most of the developed regular expressions had 1 or 2 significantly different answers.

%%new
Most of the teachers participating in developing Preg questions found creating a regular expression from a program string simple and straightforward: the best and common practice was writing the program code as it is first, then escaping special characters, then introducing alternative parts, followed by adding the notation for extra parentheses and optional whitespaces. While this process can be automated, developing a special tool for creating this question bank was found unnecessary. The regular-expression editing, visualizing, and testing interface of Preg questions were adequate to the task of developing expressions for a string of code.

While normal regular expressions often (about 11\% of times) produced false-negative grades because of extraneous parentheses in expressions, Perl-compatible regular expressions can handle these using the recursive subpattern calls feature. Practical usage of recursive subpattern calls requires special syntax for typical expressions to keep the resulting expressions human-readable; we developed several patterns using special regular-expressions comments to achieve that. After this, the amount of false-negative grades dropped to 0.8\% which can be handled by the teaching staff even in large courses. False-positive grades were extremely rare as a regular expression allows defining a set of correct answers precisely. One regular expression matched 4.9 unique students' answers on average, up to 67 unique answers matched by one expression.

%\textbf{Q2. How does voluntary usage of formative Preg quizzes affects students' performance?} We found that additional Preg formative quizzes were used mostly by medium- and low-performing students while those who can pass most of the summative quizzes from the first attempt did not use formative quizzes. However, when it comes to the final exam results, there is no significant difference between the students who used formative quizzes and those who did not. This shows that formative quizzes are a valuable tool for improving performance for students who have difficulties learning introductory programming. When failing the first attempt of a summative quiz, the learning gains for the next attempt are significantly higher (almost twice) for the students who used the formative quizzes actively preparing for the second attempt.

Analyzing attempts of the formative quizzes for training purposes, we found that when using Preg questions, the students solved at least one question correctly in almost all the attempts and solved all questions correctly in almost two-thirds of the attempts. This means that most of the students used the formative quizzes actively, answering the questions repeatedly until solution if necessary. Only about 5\% of attempts were given up without answering correctly a single question. The relatively big number of attempts where not all questions were answered correctly may be not a big issue because sometimes students omitted the questions they knew well, concentrating on the difficult questions while training. A significant number of the students' first (starting) answers contained major syntax errors (up to omitting or misplacing more than half of the answer's tokens) that would prevent a question based on error-correcting grammar from providing meaningful feedback, while Preg question could provide enough support to solve the task.

Studying question attempts, we can see that almost 80\% of them ended in a correct answer. Students used hints in about each third attempt to answer a question; the relatively low frequency of using hints can partially be attributed to the easiness of some questions. The teachers during interviewing also noted that sometimes very poorly performing students were not even aware that they can use hints. These were mostly the students who either skipped classes often or learned in the second language and had trouble understanding what was going on in the class. This last group, according to the teachers, gained a lot from using hints once they were shown how to use hints personally.

However, there is no significant difference between the percentage of attempts using hints among correctly answered and incorrectly answered attempts. In more than 13\% of cases, students gave up answering even having access to the hints that led to the correct answer if used enough times. In 73\% of the attempts when students gave up, they did not even try to use hints. This may mean that the teaching staff should spend more time teaching the students how to use hints. Still, the teachers reported a sharp decrease in the number of students' questions about formative quizzes once the hints were enabled. About 500 training questions per semester were answered correctly using hints; without hints, the students would need to ask their teacher or teaching assistant about solving the problematic question. This saves the teaching time to spend on actual programming.

%\textbf{Q3. How voluntary usage string-completion hints affects students' performance?} 
%String completion hints allowed students to continue working on their formative quizzes during their homework without being stuck. This allowed them to progress smoother and let the teaching assistants better use class time. However, asking for hints too often is consistently associated with worse results: the students who used hints more than in 1/3 attempts needed significantly more attempts of the summative quizzes to pass them, had lower learning gains from training, and lower exam quiz scores as well. An interesting finding is that the results of the students who used hints often are statistically indistinguishable from the students who did not use hints. This is likely caused by thoughtless using hints to find the correct answer.
We did not find any link between using Preg questions for training and the students' performance in the exam quiz. However, the students had two other kinds of formative quizzes to train (which the teachers found unethical to turn off) and Preg questions accounted only for open-answer questions which comprised about a third of the exam quiz. There seems to be a weak link between using hints sparingly (less than once in every 3 attempts) and performing better in the final quiz. This means that overuse of completion hints is mostly done by poorly performing students; saying simply, repeated clicks on ``show me what to type next'' does not stimulate learning. So Preg questions can be improved by adding a limiter to the number of hints the student can use. The high frequency of hint usage can serve as a warning sign about learning problems; students who do it require special attention from teaching assistants. The teachers can discourage students from using hints too often by applying grade penalties for their use as the Preg software allows.

The main positive effect of introducing regular-expression questions with string-completion hints was saving teachers' class time that was previously spent on answering students' questions about their homework quizzes. Some teachers noted that foreign students who learned in their second language and so had trouble understanding teachers gained from using Preg questions most as the string-completion hints do not require reading and understanding natural-language texts. It increased the course's inclusivity and saved a lot of teachers' time because explaining problems to these students often take a lot of time. The immediate automatic hints in code-writing quizzes worked well because the tasks are relatively simple and do not require prolonged thought. It was also very valuable during COVID lockdowns.

%So the best results were achieved by the students who used the formative quizzes actively but used hints sparingly, only to avoid being stuck in worst cases. This means that the high frequency of the hint usage and also the low number of formative quiz attempts combined with the low performance during the summative quizzes can serve as a good marker for the students who have learning problems and require special attention from teaching assistants. The teachers may also discourage students from using hints too often by applying grade penalties for their use as the Preg question allows.

\balance
\section{Limitations and Conclusion}

We found that regular expressions matching is a viable way to effectively implement short-answer questions requiring writing a single line of code (up to about 70 tokens). A single regular expression can cover up to 18 different answers. Using regular expression to grade program code requires solving the problem of extraneous parentheses in expressions that is possible using Perl-compatible regular expressions. We used the Preg question-type plug-in for Moodle LMS as a tool implementing the necessary features and able to give hints on completing partially correct students' answers. Formative quizzes with string-completion hints allow students to practice writing lines of program code performing a particular task on their own without being stuck. However, too frequent usage of hints may hinder students' progress and should be discouraged or used as a marker of poorly-performing students needing more attention. This requires further research. The main effect of training with hints seems to be saving teachers' time rather than learning gains in answering the same questions on exams: the students learned the same without consulting their teacher and teaching assistants.

These open-answer questions work only on the application level of Bloom's taxonomy of learning objectives and should be supported by other learning tools and techniques. For the assignments requiring writing more code, it is better to switch from matching strings to testing the developed code. Summative quizzes, however, stimulate mastering low-level skills before attempting high-level tasks (like solving coding assignments) and ensure a certain level of knowledge before the student attempts complex tasks requiring more attention from the teaching staff.

Further work on the Preg question type will include adding typo detection features: the most common students' complaint was about a typo causing the whole answer to be graded as wrong. Adding more features of Perl-compatible regular expressions (like complex assertions) to the hinting engine will improve its templating ability. The Preg question type is available for the widely-used Moodle LMS under General Public License and can be downloaded from %\url{https://moodle.org/plugins/qtype\_preg}.
\verb|the link is anonymised|.
\begin{acks}
The reported study was funded by RFBR, project number 20-07-00764.
\end{acks}

\bibliographystyle{ACM-Reference-Format}
\bibliography{references}


\begin{thebibliography}{34}

%%% the Web address.

\ifx \showCODEN    \undefined \def \showCODEN     #1{\unskip}     \fi
\ifx \showDOI      \undefined \def \showDOI       #1{#1}\fi
\ifx \showISBNx    \undefined \def \showISBNx     #1{\unskip}     \fi
\ifx \showISBNxiii \undefined \def \showISBNxiii  #1{\unskip}     \fi
\ifx \showISSN     \undefined \def \showISSN      #1{\unskip}     \fi
\ifx \showLCCN     \undefined \def \showLCCN      #1{\unskip}     \fi
\ifx \shownote     \undefined \def \shownote      #1{#1}          \fi
\ifx \showarticletitle \undefined \def \showarticletitle #1{#1}   \fi
\ifx \showURL      \undefined \def \showURL       {\relax}        \fi
% invisible to TeX
\providecommand\bibfield[2]{#2}
\providecommand\bibinfo[2]{#2}
\providecommand\natexlab[1]{#1}
\providecommand\showeprint[2][]{arXiv:#2}

\bibitem[\protect\citeauthoryear{Ala-Mutka}{Ala-Mutka}{2005}]%
        {AlaMutka2005}
\bibfield{author}{\bibinfo{person}{Kirsti Ala-Mutka}.}
  \bibinfo{year}{2005}\natexlab{}.
\newblock \showarticletitle{A Survey of Automated Assessment Approaches for
  Programming Assignments}.
\newblock \bibinfo{journal}{\emph{Computer Science Education}}
  \bibinfo{volume}{15} (\bibinfo{year}{2005}), \bibinfo{pages}{102 -- 83}.
\newblock
\urldef\tempurl%
\url{https://doi.org/10.1080/08993400500150747}
\showDOI{\tempurl}


\bibitem[\protect\citeauthoryear{Anderson and Krathwohl}{Anderson and
  Krathwohl}{2001}]%
        {AndersonKrathwohl2001}
\bibfield{editor}{\bibinfo{person}{Lorin~W. Anderson} {and}
  \bibinfo{person}{David~R. Krathwohl}} (Eds.).
  \bibinfo{year}{2001}\natexlab{}.
\newblock \bibinfo{booktitle}{\emph{{A Taxonomy for Learning, Teaching, and
  Assessing. A Revision of Bloom's Taxonomy of Educational Objectives}}
  (\bibinfo{edition}{2} ed.)}.
\newblock \bibinfo{publisher}{Allyn \& Bacon}, \bibinfo{address}{New York}.
\newblock
\showISBNx{978-0801319037}


\bibitem[\protect\citeauthoryear{Bachman, Carr, Kamei, Kim, Pan, Salvador, and
  Sawaki}{Bachman et~al\mbox{.}}{2002}]%
        {Bachman2002AutomaticAssessmentShortAnswer}
\bibfield{author}{\bibinfo{person}{Lyle~F Bachman}, \bibinfo{person}{Nathan
  Carr}, \bibinfo{person}{Greg Kamei}, \bibinfo{person}{Mikyung Kim},
  \bibinfo{person}{Michael~J Pan}, \bibinfo{person}{Chris Salvador}, {and}
  \bibinfo{person}{Yasuyo Sawaki}.} \bibinfo{year}{2002}\natexlab{}.
\newblock \showarticletitle{A Reliable Approach to Automatic Assessment of
  Short Answer Free Responses}. In \bibinfo{booktitle}{\emph{COLING '02:
  Proceedings of the 19th international conference on Computational
  linguistics}} (Taipei, Taiwan) \emph{(\bibinfo{series}{COLING '02})}.
  \bibinfo{publisher}{Association for Computational Linguistics},
  \bibinfo{address}{USA}, \bibinfo{pages}{1–4}.
\newblock
\urldef\tempurl%
\url{https://doi.org/10.3115/1071884.1071907}
\showDOI{\tempurl}


\bibitem[\protect\citeauthoryear{Beck, Gulan, Biegel, Baltes, and
  Weiskopf}{Beck et~al\mbox{.}}{2014}]%
        {BeckFabian2014RegViz}
\bibfield{author}{\bibinfo{person}{Fabian Beck}, \bibinfo{person}{Stefan
  Gulan}, \bibinfo{person}{Benjamin Biegel}, \bibinfo{person}{Sebastian
  Baltes}, {and} \bibinfo{person}{Daniel Weiskopf}.}
  \bibinfo{year}{2014}\natexlab{}.
\newblock \showarticletitle{RegViz: Visual Debugging of Regular Expressions}.
  In \bibinfo{booktitle}{\emph{Companion Proceedings of the 36th International
  Conference on Software Engineering}} (Hyderabad, India)
  \emph{(\bibinfo{series}{ICSE Companion 2014})}.
  \bibinfo{publisher}{Association for Computing Machinery},
  \bibinfo{address}{New York, NY, USA}, \bibinfo{pages}{504–507}.
\newblock
\showISBNx{9781450327688}
\urldef\tempurl%
\url{https://doi.org/10.1145/2591062.2591111}
\showDOI{\tempurl}


\bibitem[\protect\citeauthoryear{Black and Wiliam}{Black and Wiliam}{1998}]%
        {BlackWiliamAssessmentLearning1998}
\bibfield{author}{\bibinfo{person}{Paul Black} {and} \bibinfo{person}{Dylan
  Wiliam}.} \bibinfo{year}{1998}\natexlab{}.
\newblock \showarticletitle{Assessment and Classroom Learning}.
\newblock \bibinfo{journal}{\emph{Assessment in Education: Principles, Policy
  \& Practice}} \bibinfo{volume}{5}, \bibinfo{number}{1}
  (\bibinfo{year}{1998}), \bibinfo{pages}{7--74}.
\newblock
\urldef\tempurl%
\url{https://doi.org/10.1080/0969595980050102}
\showDOI{\tempurl}


\bibitem[\protect\citeauthoryear{{Budiselic}, {Srbljic}, and
  {Popovic}}{{Budiselic} et~al\mbox{.}}{2007}]%
        {Budiselic2007RegExpert}
\bibfield{author}{\bibinfo{person}{I. {Budiselic}}, \bibinfo{person}{S.
  {Srbljic}}, {and} \bibinfo{person}{M. {Popovic}}.}
  \bibinfo{year}{2007}\natexlab{}.
\newblock \showarticletitle{RegExpert: A Tool for Visualization of Regular
  Expressions}. In \bibinfo{booktitle}{\emph{EUROCON 2007 - The International
  Conference on "Computer as a Tool"}}. \bibinfo{publisher}{IEEE},
  \bibinfo{address}{Warsaw, Poland}, \bibinfo{pages}{2387--2389}.
\newblock
\urldef\tempurl%
\url{https://doi.org/10.1109/EURCON.2007.4400374}
\showDOI{\tempurl}


\bibitem[\protect\citeauthoryear{Burrows, Gurevych, and Stein}{Burrows
  et~al\mbox{.}}{2015}]%
        {Burrows2015ErasAutomaticAnswerGrading}
\bibfield{author}{\bibinfo{person}{Steven Burrows}, \bibinfo{person}{Iryna
  Gurevych}, {and} \bibinfo{person}{Benno Stein}.}
  \bibinfo{year}{2015}\natexlab{}.
\newblock \showarticletitle{The Eras and Trends of Automatic Short Answer
  Grading}.
\newblock \bibinfo{journal}{\emph{International Journal of Artificial
  Intelligence in Education}} \bibinfo{volume}{25}, \bibinfo{number}{1}
  (\bibinfo{date}{01 Mar} \bibinfo{year}{2015}), \bibinfo{pages}{60--117}.
\newblock
\showISSN{1560-4306}
\urldef\tempurl%
\url{https://doi.org/10.1007/s40593-014-0026-8}
\showDOI{\tempurl}


\bibitem[\protect\citeauthoryear{Butcher and Jordan}{Butcher and
  Jordan}{2010}]%
        {butcher2010comparison}
\bibfield{author}{\bibinfo{person}{Philip~G Butcher} {and}
  \bibinfo{person}{Sally~E Jordan}.} \bibinfo{year}{2010}\natexlab{}.
\newblock \showarticletitle{A comparison of human and computer marking of short
  free-text student responses}.
\newblock \bibinfo{journal}{\emph{Computers \& Education}}
  \bibinfo{volume}{55}, \bibinfo{number}{2} (\bibinfo{year}{2010}),
  \bibinfo{pages}{489--499}.
\newblock
\urldef\tempurl%
\url{https://doi.org/10.1016/j.compedu.2010.02.012}
\showDOI{\tempurl}


\bibitem[\protect\citeauthoryear{Chirumamilla and Sindre}{Chirumamilla and
  Sindre}{2019}]%
        {Chirumamilla2019EAssessment}
\bibfield{author}{\bibinfo{person}{Aparna Chirumamilla} {and}
  \bibinfo{person}{Guttorm Sindre}.} \bibinfo{year}{2019}\natexlab{}.
\newblock \showarticletitle{E-Assessment in Programming Courses: Towards a
  Digital Ecosystem Supporting Diverse Needs?}. In
  \bibinfo{booktitle}{\emph{Digital Transformation for a Sustainable Society in
  the 21st Century}}, \bibfield{editor}{\bibinfo{person}{Ilias~O. Pappas},
  \bibinfo{person}{Patrick Mikalef}, \bibinfo{person}{Yogesh~K. Dwivedi},
  \bibinfo{person}{Letizia Jaccheri}, \bibinfo{person}{John Krogstie}, {and}
  \bibinfo{person}{Matti M{\"a}ntym{\"a}ki}} (Eds.).
  \bibinfo{publisher}{Springer International Publishing},
  \bibinfo{address}{Cham}, \bibinfo{pages}{585--596}.
\newblock
\showISBNx{978-3-030-29374-1}
\urldef\tempurl%
\url{https://doi.org/10.1007/978-3-030-29374-1_47}
\showDOI{\tempurl}


\bibitem[\protect\citeauthoryear{Corbett and Anderson}{Corbett and
  Anderson}{2001}]%
        {Corbett2001Feedback}
\bibfield{author}{\bibinfo{person}{Albert~T. Corbett} {and}
  \bibinfo{person}{John~R. Anderson}.} \bibinfo{year}{2001}\natexlab{}.
\newblock \showarticletitle{Locus of Feedback Control in Computer-Based
  Tutoring: Impact on Learning Rate, Achievement and Attitudes}. In
  \bibinfo{booktitle}{\emph{Proceedings of the SIGCHI Conference on Human
  Factors in Computing Systems}} (Seattle, Washington, USA)
  \emph{(\bibinfo{series}{CHI '01})}. \bibinfo{publisher}{Association for
  Computing Machinery}, \bibinfo{address}{New York, NY, USA},
  \bibinfo{pages}{245–252}.
\newblock
\showISBNx{1581133278}
\urldef\tempurl%
\url{https://doi.org/10.1145/365024.365111}
\showDOI{\tempurl}


\bibitem[\protect\citeauthoryear{{Dezso Sima}, {Balazs Schmuck}, {Sandor
  Szollosi}, and {Arpad Miklos}}{{Dezso Sima} et~al\mbox{.}}{2007}]%
        {Sima2007IntelligentShortTextAssessment}
\bibfield{author}{\bibinfo{person}{{Dezso Sima}}, \bibinfo{person}{{Balazs
  Schmuck}}, \bibinfo{person}{{Sandor Szollosi}}, {and} \bibinfo{person}{{Arpad
  Miklos}}.} \bibinfo{year}{2007}\natexlab{}.
\newblock \showarticletitle{Intelligent short text assessment in eMax}. In
  \bibinfo{booktitle}{\emph{AFRICON 2007}}. \bibinfo{publisher}{IEEE},
  \bibinfo{address}{Windhoek}, \bibinfo{pages}{1--7}.
\newblock
\urldef\tempurl%
\url{https://doi.org/10.1109/AFRCON.2007.4401593}
\showDOI{\tempurl}


\bibitem[\protect\citeauthoryear{Dzikovska, Steinhauser, Farrow, Moore, and
  Campbell}{Dzikovska et~al\mbox{.}}{2014}]%
        {Dzikovska2014}
\bibfield{author}{\bibinfo{person}{Myroslava Dzikovska},
  \bibinfo{person}{Natalie Steinhauser}, \bibinfo{person}{Elaine Farrow},
  \bibinfo{person}{Johanna Moore}, {and} \bibinfo{person}{Gwendolyn Campbell}.}
  \bibinfo{year}{2014}\natexlab{}.
\newblock \showarticletitle{BEETLE II: Deep Natural Language Understanding and
  Automatic Feedback Generation for Intelligent Tutoring in Basic Electricity
  and Electronics}.
\newblock \bibinfo{journal}{\emph{International Journal of Artificial
  Intelligence in Education}} \bibinfo{volume}{24}, \bibinfo{number}{3}
  (\bibinfo{date}{01 Sep} \bibinfo{year}{2014}), \bibinfo{pages}{284--332}.
\newblock
\showISSN{1560-4306}
\urldef\tempurl%
\url{https://doi.org/10.1007/s40593-014-0017-9}
\showDOI{\tempurl}


\bibitem[\protect\citeauthoryear{Edwards, Murali, and Kazerouni}{Edwards
  et~al\mbox{.}}{2019}]%
        {Edwards2019}
\bibfield{author}{\bibinfo{person}{Stephen~H. Edwards},
  \bibinfo{person}{Krishnan~P. Murali}, {and} \bibinfo{person}{Ayaan~M.
  Kazerouni}.} \bibinfo{year}{2019}\natexlab{}.
\newblock \showarticletitle{The Relationship Between Voluntary Practice of
  Short Programming Exercises and Exam Performance}. In
  \bibinfo{booktitle}{\emph{Proceedings of the ACM Conference on Global
  Computing Education}} (Chengdu,Sichuan, China) \emph{(\bibinfo{series}{CompEd
  '19})}. \bibinfo{publisher}{Association for Computing Machinery},
  \bibinfo{address}{New York, NY, USA}, \bibinfo{pages}{113–119}.
\newblock
\showISBNx{9781450362597}
\urldef\tempurl%
\url{https://doi.org/10.1145/3300115.3309525}
\showDOI{\tempurl}


\bibitem[\protect\citeauthoryear{Estey and Coady}{Estey and Coady}{2017}]%
        {Estey2017}
\bibfield{author}{\bibinfo{person}{Anthony Estey} {and} \bibinfo{person}{Yvonne
  Coady}.} \bibinfo{year}{2017}\natexlab{}.
\newblock \showarticletitle{Study Habits, Exam Performance, and Confidence: How
  Do Workflow Practices and Self-Efficacy Ratings Align?}. In
  \bibinfo{booktitle}{\emph{Proceedings of the 2017 ACM Conference on
  Innovation and Technology in Computer Science Education}} (Bologna, Italy)
  \emph{(\bibinfo{series}{ITiCSE '17})}. \bibinfo{publisher}{Association for
  Computing Machinery}, \bibinfo{address}{New York, NY, USA},
  \bibinfo{pages}{158–163}.
\newblock
\showISBNx{9781450347044}
\urldef\tempurl%
\url{https://doi.org/10.1145/3059009.3059056}
\showDOI{\tempurl}


\bibitem[\protect\citeauthoryear{Falkner, Vivian, Piper, and Falkner}{Falkner
  et~al\mbox{.}}{2014}]%
        {Falkner2014}
\bibfield{author}{\bibinfo{person}{Nickolas Falkner}, \bibinfo{person}{Rebecca
  Vivian}, \bibinfo{person}{David Piper}, {and} \bibinfo{person}{Katrina
  Falkner}.} \bibinfo{year}{2014}\natexlab{}.
\newblock \showarticletitle{Increasing the Effectiveness of Automated
  Assessment by Increasing Marking Granularity and Feedback Units}. In
  \bibinfo{booktitle}{\emph{Proceedings of the 45th ACM Technical Symposium on
  Computer Science Education}} (Atlanta, Georgia, USA)
  \emph{(\bibinfo{series}{SIGCSE '14})}. \bibinfo{publisher}{Association for
  Computing Machinery}, \bibinfo{address}{New York, NY, USA},
  \bibinfo{pages}{9–14}.
\newblock
\showISBNx{9781450326056}
\urldef\tempurl%
\url{https://doi.org/10.1145/2538862.2538896}
\showDOI{\tempurl}


\bibitem[\protect\citeauthoryear{Friedl}{Friedl}{2006}]%
        {friedl2006mastering}
\bibfield{author}{\bibinfo{person}{Jeffrey E.~F. Friedl}.}
  \bibinfo{year}{2006}\natexlab{}.
\newblock \bibinfo{booktitle}{\emph{Mastering Regular Expressions}
  (\bibinfo{edition}{3} ed.)}.
\newblock \bibinfo{publisher}{O'Reilly}, \bibinfo{address}{Beijing}.
\newblock
\showISBNx{978-0-596-52812-6}


\bibitem[\protect\citeauthoryear{Gerdes, Jeuring, and Heeren}{Gerdes
  et~al\mbox{.}}{2012}]%
        {Gerdes2012FunctionalTutor}
\bibfield{author}{\bibinfo{person}{Alex Gerdes}, \bibinfo{person}{Johan
  Jeuring}, {and} \bibinfo{person}{Bastiaan Heeren}.}
  \bibinfo{year}{2012}\natexlab{}.
\newblock \showarticletitle{An Interactive Functional Programming Tutor}. In
  \bibinfo{booktitle}{\emph{Annual Conference on Innovation and Technology in
  Computer Science Education, ITiCSE}}. \bibinfo{publisher}{Association for
  Computing Machinery}, \bibinfo{address}{New York, NY, USA}.
\newblock
\urldef\tempurl%
\url{https://doi.org/10.1145/2325296.2325356}
\showDOI{\tempurl}


\bibitem[\protect\citeauthoryear{Gibbs and Simpson}{Gibbs and Simpson}{2005}]%
        {gibbs2005conditions}
\bibfield{author}{\bibinfo{person}{Graham Gibbs} {and} \bibinfo{person}{Claire
  Simpson}.} \bibinfo{year}{2005}\natexlab{}.
\newblock \showarticletitle{Conditions under which assessment supports
  students’ learning}.
\newblock \bibinfo{journal}{\emph{Learning and teaching in higher education}}
  \bibinfo{volume}{1} (\bibinfo{year}{2005}), \bibinfo{pages}{3--31}.
\newblock
\urldef\tempurl%
\url{http://eprints.glos.ac.uk/3609/}
\showURL{%
\tempurl}


\bibitem[\protect\citeauthoryear{Goyvaerts and Levithan}{Goyvaerts and
  Levithan}{2012}]%
        {goyvaerts2012regular}
\bibfield{author}{\bibinfo{person}{Jan Goyvaerts} {and} \bibinfo{person}{Steven
  Levithan}.} \bibinfo{year}{2012}\natexlab{}.
\newblock \bibinfo{booktitle}{\emph{Regular expressions cookbook}}.
\newblock \bibinfo{publisher}{O'reilly}, \bibinfo{address}{USA, Sebastopol}.
\newblock


\bibitem[\protect\citeauthoryear{Hattie and Timperley}{Hattie and
  Timperley}{2007}]%
        {Hattie2007PowerFeedback}
\bibfield{author}{\bibinfo{person}{John Hattie} {and} \bibinfo{person}{Helen
  Timperley}.} \bibinfo{year}{2007}\natexlab{}.
\newblock \showarticletitle{The Power of Feedback}.
\newblock \bibinfo{journal}{\emph{Review of Educational Research}}
  \bibinfo{volume}{77} (\bibinfo{date}{03} \bibinfo{year}{2007}),
  \bibinfo{pages}{81--112}.
\newblock
\urldef\tempurl%
\url{https://doi.org/10.3102/003465430298487}
\showDOI{\tempurl}


\bibitem[\protect\citeauthoryear{Higgins, Gray, Symeonidis, and
  Tsintsifas}{Higgins et~al\mbox{.}}{2005}]%
        {Higgins2005}
\bibfield{author}{\bibinfo{person}{Colin~A. Higgins}, \bibinfo{person}{Geoffrey
  Gray}, \bibinfo{person}{Pavlos Symeonidis}, {and} \bibinfo{person}{Athanasios
  Tsintsifas}.} \bibinfo{year}{2005}\natexlab{}.
\newblock \showarticletitle{Automated Assessment and Experiences of Teaching
  Programming}.
\newblock \bibinfo{journal}{\emph{J. Educ. Resour. Comput.}}
  \bibinfo{volume}{5}, \bibinfo{number}{3} (\bibinfo{date}{Sept.}
  \bibinfo{year}{2005}), \bibinfo{pages}{5–es}.
\newblock
\showISSN{1531-4278}
\urldef\tempurl%
\url{https://doi.org/10.1145/1163405.1163410}
\showDOI{\tempurl}


\bibitem[\protect\citeauthoryear{Jordan and Mitchell}{Jordan and
  Mitchell}{2009}]%
        {Jordan2009PotentialShortAnswer}
\bibfield{author}{\bibinfo{person}{Sally Jordan} {and} \bibinfo{person}{Tom
  Mitchell}.} \bibinfo{year}{2009}\natexlab{}.
\newblock \showarticletitle{e-Assessment for learning? The potential of
  short-answer free-text questions with tailored feedback}.
\newblock \bibinfo{journal}{\emph{British Journal of Educational Technology}}
  \bibinfo{volume}{40}, \bibinfo{number}{2} (\bibinfo{year}{2009}),
  \bibinfo{pages}{371--385}.
\newblock
\urldef\tempurl%
\url{https://doi.org/10.1111/j.1467-8535.2008.00928.x}
\showDOI{\tempurl}


\bibitem[\protect\citeauthoryear{Keuning, Jeuring, and Heeren}{Keuning
  et~al\mbox{.}}{2016}]%
        {Keuning2016SystematicReviewAutomatedFeedback}
\bibfield{author}{\bibinfo{person}{Hieke Keuning}, \bibinfo{person}{Johan
  Jeuring}, {and} \bibinfo{person}{Bastiaan Heeren}.}
  \bibinfo{year}{2016}\natexlab{}.
\newblock \bibinfo{booktitle}{\emph{Towards a Systematic Review of Automated
  Feedback Generation for Programming Exercises}}.
\newblock \bibinfo{publisher}{Association for Computing Machinery},
  \bibinfo{address}{New York, NY, USA}, \bibinfo{pages}{41–46}.
\newblock
\showISBNx{9781450342315}
\urldef\tempurl%
\url{https://doi.org/10.1145/2899415.2899422}
\showURL{%
\tempurl}


\bibitem[\protect\citeauthoryear{Keuning, Jeuring, and Heeren}{Keuning
  et~al\mbox{.}}{2018}]%
        {Keuning2018FeedbackProgrammingExercises}
\bibfield{author}{\bibinfo{person}{Hieke Keuning}, \bibinfo{person}{Johan
  Jeuring}, {and} \bibinfo{person}{Bastiaan Heeren}.}
  \bibinfo{year}{2018}\natexlab{}.
\newblock \showarticletitle{A Systematic Literature Review of Automated
  Feedback Generation for Programming Exercises}.
\newblock \bibinfo{journal}{\emph{ACM Trans. Comput. Educ.}}
  \bibinfo{volume}{19}, \bibinfo{number}{1}, Article \bibinfo{articleno}{3}
  (\bibinfo{date}{Sept.} \bibinfo{year}{2018}), \bibinfo{numpages}{43}~pages.
\newblock
\urldef\tempurl%
\url{https://doi.org/10.1145/3231711}
\showDOI{\tempurl}


\bibitem[\protect\citeauthoryear{Lane and Vanlehn}{Lane and Vanlehn}{2005}]%
        {Lane2005NLTutoring}
\bibfield{author}{\bibinfo{person}{H. Lane} {and} \bibinfo{person}{Kurt
  Vanlehn}.} \bibinfo{year}{2005}\natexlab{}.
\newblock \showarticletitle{Teaching the tacit knowledge of programming to
  noviceswith natural language tutoring}.
\newblock \bibinfo{journal}{\emph{Computer Science Education}}
  \bibinfo{volume}{15} (\bibinfo{date}{09} \bibinfo{year}{2005}),
  \bibinfo{pages}{183--201}.
\newblock
\urldef\tempurl%
\url{https://doi.org/10.1080/08993400500224286}
\showDOI{\tempurl}


\bibitem[\protect\citeauthoryear{Le and Menzel}{Le and Menzel}{2006}]%
        {Nguyen2006LogicProgramming}
\bibfield{author}{\bibinfo{person}{Nguyen-Thinh Le} {and}
  \bibinfo{person}{Wolfgang Menzel}.} \bibinfo{year}{2006}\natexlab{}.
\newblock \showarticletitle{Problem Solving Process Oriented Diagnosis in Logic
  Programming}. In \bibinfo{booktitle}{\emph{Proceedings of the 2006 Conference
  on Learning by Effective Utilization of Technologies: Facilitating
  Intercultural Understanding}}. \bibinfo{publisher}{IOS Press},
  \bibinfo{address}{NLD}, \bibinfo{pages}{63–70}.
\newblock
\showISBNx{1586036874}
\urldef\tempurl%
\url{https://doi.org/10.5555/1565941.1565955}
\showDOI{\tempurl}


\bibitem[\protect\citeauthoryear{Luxton-Reilly, Simon, Albluwi, Becker,
  Giannakos, Kumar, Ott, Paterson, Scott, Sheard, and Szabo}{Luxton-Reilly
  et~al\mbox{.}}{2018}]%
        {Luxton2018}
\bibfield{author}{\bibinfo{person}{Andrew Luxton-Reilly},
  \bibinfo{person}{Simon}, \bibinfo{person}{Ibrahim Albluwi},
  \bibinfo{person}{Brett~A. Becker}, \bibinfo{person}{Michail Giannakos},
  \bibinfo{person}{Amruth~N. Kumar}, \bibinfo{person}{Linda Ott},
  \bibinfo{person}{James Paterson}, \bibinfo{person}{Michael~James Scott},
  \bibinfo{person}{Judy Sheard}, {and} \bibinfo{person}{Claudia Szabo}.}
  \bibinfo{year}{2018}\natexlab{}.
\newblock \showarticletitle{Introductory Programming: A Systematic Literature
  Review}. In \bibinfo{booktitle}{\emph{Proceedings Companion of the 23rd
  Annual ACM Conference on Innovation and Technology in Computer Science
  Education}} (Larnaca, Cyprus) \emph{(\bibinfo{series}{ITiCSE 2018
  Companion})}. \bibinfo{publisher}{Association for Computing Machinery},
  \bibinfo{address}{New York, NY, USA}, \bibinfo{pages}{55–--106}.
\newblock
\showISBNx{9781450362238}
\urldef\tempurl%
\url{https://doi.org/10.1145/3293881.3295779}
\showDOI{\tempurl}


\bibitem[\protect\citeauthoryear{Macnish}{Macnish}{2010}]%
        {Macnish2010AutomatingFeedback}
\bibfield{author}{\bibinfo{person}{Cara Macnish}.}
  \bibinfo{year}{2010}\natexlab{}.
\newblock \showarticletitle{Java Facilities for Automating Analysis, Feedback
  and Assessment of Laboratory Work}.
\newblock \bibinfo{journal}{\emph{Computer Science Education}}
  \bibinfo{volume}{August 2000} (\bibinfo{date}{08} \bibinfo{year}{2010}),
  \bibinfo{pages}{147--163}.
\newblock
\urldef\tempurl%
\url{https://doi.org/10.1076/0899-3408(200008)10:2;1-C;FT147}
\showDOI{\tempurl}


\bibitem[\protect\citeauthoryear{Marin, Contractor, and Rivero}{Marin
  et~al\mbox{.}}{2021}]%
        {Marin2021}
\bibfield{author}{\bibinfo{person}{Victor~J. Marin},
  \bibinfo{person}{Maheen~Riaz Contractor}, {and} \bibinfo{person}{Carlos~R.
  Rivero}.} \bibinfo{year}{2021}\natexlab{}.
\newblock \showarticletitle{Flexible Program Alignment to Deliver Data-Driven
  Feedback to Novice Programmers}. In \bibinfo{booktitle}{\emph{Intelligent
  Tutoring Systems}}, \bibfield{editor}{\bibinfo{person}{Alexandra~I. Cristea}
  {and} \bibinfo{person}{Christos Troussas}} (Eds.).
  \bibinfo{publisher}{Springer International Publishing},
  \bibinfo{address}{Cham}, \bibinfo{pages}{247--258}.
\newblock
\showISBNx{978-3-030-80421-3}
\urldef\tempurl%
\url{https://doi.org/10.1007/978-3-030-80421-3_27}
\showDOI{\tempurl}


\bibitem[\protect\citeauthoryear{Ott, Robins, and Shephard}{Ott
  et~al\mbox{.}}{2016}]%
        {Ott2016}
\bibfield{author}{\bibinfo{person}{Claudia Ott}, \bibinfo{person}{Anthony
  Robins}, {and} \bibinfo{person}{Kerry Shephard}.}
  \bibinfo{year}{2016}\natexlab{}.
\newblock \showarticletitle{Translating Principles of Effective Feedback for
  Students into the CS1 Context}.
\newblock \bibinfo{journal}{\emph{ACM Trans. Comput. Educ.}}
  \bibinfo{volume}{16}, \bibinfo{number}{1}, Article \bibinfo{articleno}{1}
  (\bibinfo{date}{Jan.} \bibinfo{year}{2016}), \bibinfo{numpages}{27}~pages.
\newblock
\urldef\tempurl%
\url{https://doi.org/10.1145/2737596}
\showDOI{\tempurl}


\bibitem[\protect\citeauthoryear{Rajasekaran and Nicolae}{Rajasekaran and
  Nicolae}{2014}]%
        {rajasekaran2014error}
\bibfield{author}{\bibinfo{person}{Sanguthevar Rajasekaran} {and}
  \bibinfo{person}{Marius Nicolae}.} \bibinfo{year}{2014}\natexlab{}.
\newblock \bibinfo{title}{An error correcting parser for context free grammars
  that takes less than cubic time}.
\newblock
\newblock
\showeprint[arxiv]{1406.3405}~[cs.DS]


\bibitem[\protect\citeauthoryear{{Siddiqi} and {Harrison}}{{Siddiqi} and
  {Harrison}}{2008}]%
        {Siddiqi2008AutomatedMarkingShA}
\bibfield{author}{\bibinfo{person}{R. {Siddiqi}} {and} \bibinfo{person}{C.
  {Harrison}}.} \bibinfo{year}{2008}\natexlab{}.
\newblock \showarticletitle{A systematic approach to the automated marking of
  short-answer questions}. In \bibinfo{booktitle}{\emph{2008 IEEE International
  Multitopic Conference}}. \bibinfo{publisher}{IEEE},
  \bibinfo{address}{Karachi}, \bibinfo{pages}{329--332}.
\newblock
\urldef\tempurl%
\url{https://doi.org/10.1109/INMIC.2008.4777758}
\showDOI{\tempurl}


\bibitem[\protect\citeauthoryear{Spacco, Denny, Richards, Babcock, Hovemeyer,
  Moscola, and Duvall}{Spacco et~al\mbox{.}}{2015}]%
        {Spacco2015}
\bibfield{author}{\bibinfo{person}{Jaime Spacco}, \bibinfo{person}{Paul Denny},
  \bibinfo{person}{Brad Richards}, \bibinfo{person}{David Babcock},
  \bibinfo{person}{David Hovemeyer}, \bibinfo{person}{James Moscola}, {and}
  \bibinfo{person}{Robert Duvall}.} \bibinfo{year}{2015}\natexlab{}.
\newblock \showarticletitle{Analyzing Student Work Patterns Using Programming
  Exercise Data}. In \bibinfo{booktitle}{\emph{Proceedings of the 46th ACM
  Technical Symposium on Computer Science Education}} (Kansas City, Missouri,
  USA) \emph{(\bibinfo{series}{SIGCSE '15})}. \bibinfo{publisher}{Association
  for Computing Machinery}, \bibinfo{address}{New York, NY, USA},
  \bibinfo{pages}{18–23}.
\newblock
\showISBNx{9781450329668}
\urldef\tempurl%
\url{https://doi.org/10.1145/2676723.2677297}
\showDOI{\tempurl}


\bibitem[\protect\citeauthoryear{Sychev, Anikin, and Prokudin}{Sychev
  et~al\mbox{.}}{2020}]%
        {SYCHEV2020264}
\bibfield{author}{\bibinfo{person}{Oleg Sychev}, \bibinfo{person}{Anton
  Anikin}, {and} \bibinfo{person}{Artem Prokudin}.}
  \bibinfo{year}{2020}\natexlab{}.
\newblock \showarticletitle{Automatic grading and hinting in open-ended text
  questions}.
\newblock \bibinfo{journal}{\emph{Cognitive Systems Research}}
  \bibinfo{volume}{59} (\bibinfo{year}{2020}), \bibinfo{pages}{264 -- 272}.
\newblock
\showISSN{1389-0417}
\urldef\tempurl%
\url{https://doi.org/10.1016/j.cogsys.2019.09.025}
\showDOI{\tempurl}


\end{thebibliography}


\end{document}